\documentclass[twocolumn]{aa}  

\usepackage{graphicx}
\usepackage{multirow}
\usepackage{txfonts}
\usepackage{subfigure}
\usepackage{color}
\usepackage[switch]{lineno}

\begin{document} 

   \title{Realization of a multifrequency celestial reference frame through a combination of normal equation systems}
    \titlerunning{Multifrequency CRF through combination}

   \author{M. Karbon \inst{1, 2}
          \and
          A. Nothnagel \inst{1}
          }

   \institute{Institute of Geodesy and Geoinformation, University of Bonn,
              Nussallee 17, D-53115 Bonn, Germany \and
              Current address: Observatoire de Paris, D\'{e}partement Syst\`{e}mes de R\'{e}f\'{e}rence Temps Espace (SYRTE),
              75014 Paris, France\\
              \email{maria.karbon@obspm.fr}
             }

   \date{Received ; accepted }

 
  \abstract
   {We present a celestial reference frame (CRF) based on the combination of independent, multifrequency radio source position catalogs using nearly 40 years of Very Long Baseline Interferometry observations at the standard geodetic frequencies at SX band and about 15 years of observations at higher frequencies (K and XKa). The final catalog contains 4617 sources.}
   {We produce a multifrequency catalog of radio source positions with full variance-covariance information across all radio source positions of all input catalogs.}
   {We combined three catalogs, one observed at 8 GHz (X band), one at 24 GHz (K band) and one at 32 GHz (Ka band). Rather than only using the radio source positions, we developed a new, rigorous combination approach  by carrying over the full covariance information through the process of adding normal equation systems. Special validation routines were used to characterize the random and systematic errors between the input reference frames and the combined catalog.}
   {The resulting  CRF contains precise positions of 4617 compact radio astronomical objects, 4536 measured at 8~Ghz, 824 sources also observed at 24 GHz, and 674 at 32 GHz. The frame is aligned with ICRF3 within $\pm$3 $\mu$as and shows an average positional uncertainty of 0.1 mas in right ascension and declination. No significant deformations can be identified. Comparisons with Gaia-CRF remain inconclusive, nonetheless significant differences between all frames can be attested.   }
   {}


   \maketitle
%

\section{Introduction}
The third realization of the International Celestial Reference System (ICRS), the International Celestial Reference Frame \#3 (ICRF3) based on VLBI observations at radio frequencies was adopted by the International Astronomical Union (IAU) in August 2018 \citep{Charlot2019}. The ICRF3 features source positions predominantly in the X band frequency (8 GHz). However in addition, two independent catalogs observed at higher frequencies at K (22~GHz) and Ka band (32~GHz), are included as well. The three input catalogs composing ICRF3 are all computed through individual monolithic solutions. These catalogs are aligned to the X band catalog through a match of common candidates for the transfer of the datum \citep{Charlot2019}. 

In this paper, we present the combination of independent, multifrequency radio source position catalogs in a different manner than that used for ICRF3. Our new approach combines these catalogs by carrying over the full covariance information of each catalog through the process of an accumulation of normal equation systems instead of using only the positions themselves. Through this novel process, a complete covariance matrix of the entire set of sources across the three bands is provided.

We are aware that there are issues related to core shift \citep{Plavin2019b}, which may cause the matching procedures of identical catalogs at different frequency bands to become insufficient for a variety of candidates. However, in this publication, we demonstrate that the effect is of a random nature for catalog combinations. For future developments, our procedure for rigorous catalog combination can easily be expanded when reliable core shift information becomes available for a larger number of sources.

The combination of position catalogs has a long history. Without going even further back in time, fundamental star catalogs such as FK4 and FK5  \cite[e.g.,~][]{Fricke1963,Fricke1988} were produced in a compilation process and represented important conventional celestial reference frames (CRF). With the advent of high-precision geodetic VLBI in the late 1970s, the positions of astronomical objects in the radio frequency domain gained importance. Catalog combinations and comparisons by interpreting post-fit residuals  \citep[e.g.,~][]{Brosche1966,Brosche1970,Eichhorn1974} or using the arc-length method \cite[e.g.,~][]{Yatskiv1990} were important investigations in that era. Later, the work on optical catalogs was revived through the Hipparcos mission \citep{Perryman1997}. Various combination efforts linked the Hipparcos catalog to other optical frames or to radio reference frames \cite[e.g.,~][]{Lestrade1995}.

From 1997 onward, the IAU adopted position catalogs based on observations in the radio frequency domain as the official ICRF. Their first versions, ICRF \citep{Ma1998} and ICRF2 \citep{ICRF2}, were both computed from a single monolithic solution of all geodetic and astrometric VLBI observations at X band gathered since the middle of 1979. X band around 8.4 GHz is the primary observing frequency while S band observations near 2.3 GHz serve solely for ionospheric calibrations, thus, the commonly used denomination SX (and analogously XKa). The third realization, ICRF3 \citep{Charlot2019}, is a multifrequency catalog consisting of three individual catalogs containing the source coordinates observed at different frequencies. The SX solution was again computed from one single monolithic VLBI solution, and the two catalogs at higher frequencies (K and XKa band) were aligned with it.

Geodetic and astrometric VLBI data analysis consists of a general two-step procedure. First, each new session data set of 24 h duration with its group delay observables is preprocessed in terms of calibrations, parameterization, and outlier elimination. By fixing certain frame parameters, the first preliminary results can be deduced. The main results are, however, derived from so-called global solutions, which use as input all preprocessed observing sessions accumulated so far. Parameters to be estimated in these solutions include radio telescope coordinates with their time derivatives reflecting tectonic motion, Earth orientation parameters, and radio source positions. Because these positions are an integral part of the parameter list, geodetic and astrometric solutions cannot be separated in concept. Solutions can, however, be different in terms of datum definition, both for the terrestrial and celestial aspect. Up-to-date VLBI solutions are based on current geophysical models following the agreed-upon standard of the International Earth Rotation and Reference Systems Service (IERS), the IERS Conventions 2010 \citep{IERS2010}.

Concerning the combination of individual solutions into a consistent composition with a transfer of the full variance-covariance information (rigorous combination), \citet{Iddink2015} first demonstrated the feasibility of such a process chain. \citet{Bachmann2017} applied this method to session-wise data sets of the International VLBI Service for Geodesy and Astrometry (IVS) and confirmed the validity of the concept, although these authors had a different aim. For the first two realizations of the ICRF and ICRF2, such developments and software tools had not been available and they had been computed only as monolithic single frequency (X band) solutions from a single analysis center with a single analysis software. 

Although ICRF3 also contains source positions at higher frequencies (i.e., K and Ka band), the respective catalogs are stand-alone products, and are only aligned to the ICRS via identical sources in the individual catalogs pertaining to the so-called ICRF3 defining sources. A rigorous combination, however, can improve the robustness and stability of the final product by mitigating smaller error effects and systematics inherent to a single solution \citep{Beutler1995}. \citet{Boeckmann2010_EOPcombo} have clearly shown the advantages of a combination of Earth orientation parameters in VLBI, which results in a more stable and robust solution. The same is valid for the determination of the international terrestrial reference frame \citep[ITRF,][]{Altamimi2007}, so that combining the results from different analysis centers has become standard for the determination of most of the products of the IERS \citep{IERS2017_annualReport}. The exception is the ICRF, as the existing solutions of the SX observations of other IVS analysis centers could not contribute to the ICRF3 for a lack of readiness of data handling and the combination process for this purpose. Comparisons have only been made between individual solutions for an assessment of the level of agreement \citep{Charlot2019}.

In our publication, we describe the input data (Sec. \ref{ch:data}) and discuss their quality (Sec. \ref{ch:quicklook}) followed by Sec. \ref{ch:angular} substantiating the validity of neglecting core shift effects. We then come to the core of this publication and present the mechanism applied for the combination (Sec. \ref{ch:method}). We characterize our results in Sec. \ref{ch:results} and close with conclusions and outlook  in Sec. \ref{ch:conclusions}.

\section{Data}
\label{ch:data}
\subsection{Input solutions}
The data in our studies are neither raw VLBI observables nor solutions of these observables in the form of position catalogs. We rather use pre-reduced normal equation systems (cf. Sec. \ref{ch:method}) of the least-squares adjustment processes that are generated on the way to the catalog solutions. These systems only contain normal equation elements that refer to the source positions, while all other parameters such as Earth orientation parameters, station positions, and the like are pre-reduced (eliminated) \citep{Boeckmann2010_VTRF}. These matrices contain all the information necessary to carry over the variance-covariance relationships of the parameters to be estimated. For the sake of generality, we still call each input data set a solution or a catalog. The data were made available to us primarily in the so-called SINEX format (\textbf{S}olution \textbf{IN}dependent \textbf{EX}change format), which reports \textit{a priori} positions, normal equations, right hand vector (b-vector), and some statistical figures such as number of observations and unknowns.

The overview of the individual contributions to the combination concentrates on a few important aspects of each solution while the procedures of their generation are largely  described in \cite{Charlot2019} and the references provided within. In total, five solutions were submitted: one each from the VLBI analysis groups of NASA Goddard Space Flight Center (GSF), German Research Centre for Geosciences (GFZ), and Vienna University of Technology (VIE) resulting from SX observations; one from the NASA Jet Propulsion Laboratory from XKa observations (XKa); and another from the NASA GSF from observations at K band (K). In terms of data content, one exception exists. The XKa solution was not provided as a normal equation system but as a catalog with its full covariance matrix because a square root information filter was used in the inversion of the observations and thus normal equations are not generated. 
For the first time a galactic aberration model was applied during the generation of all catalogs for epoch 2015.0 using a galactic aberration constant of 5.8 $\mu$as/yr \citep{Charlot2019}. 

\subsubsection{SX band} 
The catalogs in the SX frequency bands are based on almost the entire archive of VLBI sessions observed and made available by the IVS \citep{Nothnagel2017}. The observation time spans from August 1979 to March 2018 and encompasses a total of 6206 sessions, typically of 24 h duration. This period of time is required to estimate parameters for nutation and polar motion reliably and to average out the remaining unmodeled geophysical effects.
In contrast to previous ICRS realizations, all sources with more than three observations  were estimated as global parameters, meaning that all sources are represented by a single position assumed to be valid for the whole observing period. Gravitational lenses, stars, and other unsuitable objects were excluded. The celestial datum was realized through a no-net-rotation (NNR) condition \citep{JACOBS:2010} on the 295 ICRF2 defining sources. 
Because of the link between the celestial and terrestrial reference frame, station positions and linear velocities caused by tectonic motions such as continental drift were estimated in the solutions as well. The terrestrial datum was realized though NNR and no-net-translation conditions on a set of selected radio telescopes not exhibiting breaks or jumps in their positions. All other parameters are estimated as arc parameters, and therefore they are only valid for the respective observing session.

\begin{itemize}


\item GSF: The catalog submitted by the VLBI group at NASA GSF  was determined using the Calc/Solve software package \citep{Ma1990}. It has a long development history starting in the late 1970s. The SINEX file contains the positions of 4536 sources and the corresponding normal equation system and the right-hand side (b-vector). This solution is identical in content to that used for the determination of the SX catalog of ICRF3 \citep{Charlot2019}.

\item GFZ: The VLBI analysis group at the GFZ German Research Centre for Geosciences uses the VieVS$@$GFZ software \citep{Nilsson2015}, a derivative of the Vienna VLBI software VieVS \citep{Boehm_2018}. The major difference is that it allows the estimation of all parameters of interest using a Kalman filter. However, for the estimation of the CRF, this feature was not used; instead the standard method, that is the least-squares adjustment, was applied. The submitted SINEX file contains the information for 4537 sources.

\item VIE: The VieVS catalog has been developed and maintained since the early 2000s at the Vienna University of Technology. The submitted SINEX file contains information for 4521 sources.
\end{itemize}

The different number of sources contained in the catalogs originates in different analysis settings applied by the individual analysts. Different observations might be flagged as outliers or entire sessions may be excluded from the solution owing to poor performance. This can lead to an insufficient total number of observations for some sources, impeding the estimations of the coordinates.

\subsubsection{K band}
The catalog resulting from observations at K band is dominated by Very Long Baseline Array (VLBA) observations. In total, 40 sessions were observed by the VLBA between May 2002 and May 2018. However, as this array can only observe down to mid-southern declinations, additional observations on the baseline Hartebeesthoek (South Africa) to Hobart (Australia) were scheduled providing data of 16 single-baseline sessions between 2014 and 2018. As K band observations are performed with a single frequency only, the ionospheric corrections were derived from maps of total electron content generated from daily worldwide global positioning system (GPS) observations. The Calc/Solve software was used for the analysis. The final catalog comprises the normal equations for 824 sources that formed the basis for the ICRF3 K band solution \citep{Charlot2019}.

\subsubsection{XKa band} 
The observations in the XKa frequency bands were carried out predominantly by the  Deep Space Network (DSN) with stations in Goldstone (California), Robledo (Spain), and Tidbinbilla (Australia) \citep{Jacobs2012}. The observations started in 2005 with the aim to build a reference frame for deep space navigation, and 167 single baseline sessions have been observed so far. To strengthen the geometry and reduce systematics in the reference frame, the European Space Agency (ESA) telescope in Malargue in Argentina was employed for 10$\%$ of these sessions. The observations were analyzed using the MODEST software \citep{Sovers1998} applying a square root information filter for the parameter estimation. The catalog we used contains 678 sources and predates that used for the ICRF3 \citep{Charlot2019} by six months.

\subsection{Unique sources}
\begin{figure}[h!]
\centering
   \resizebox{\hsize}{!} {\includegraphics{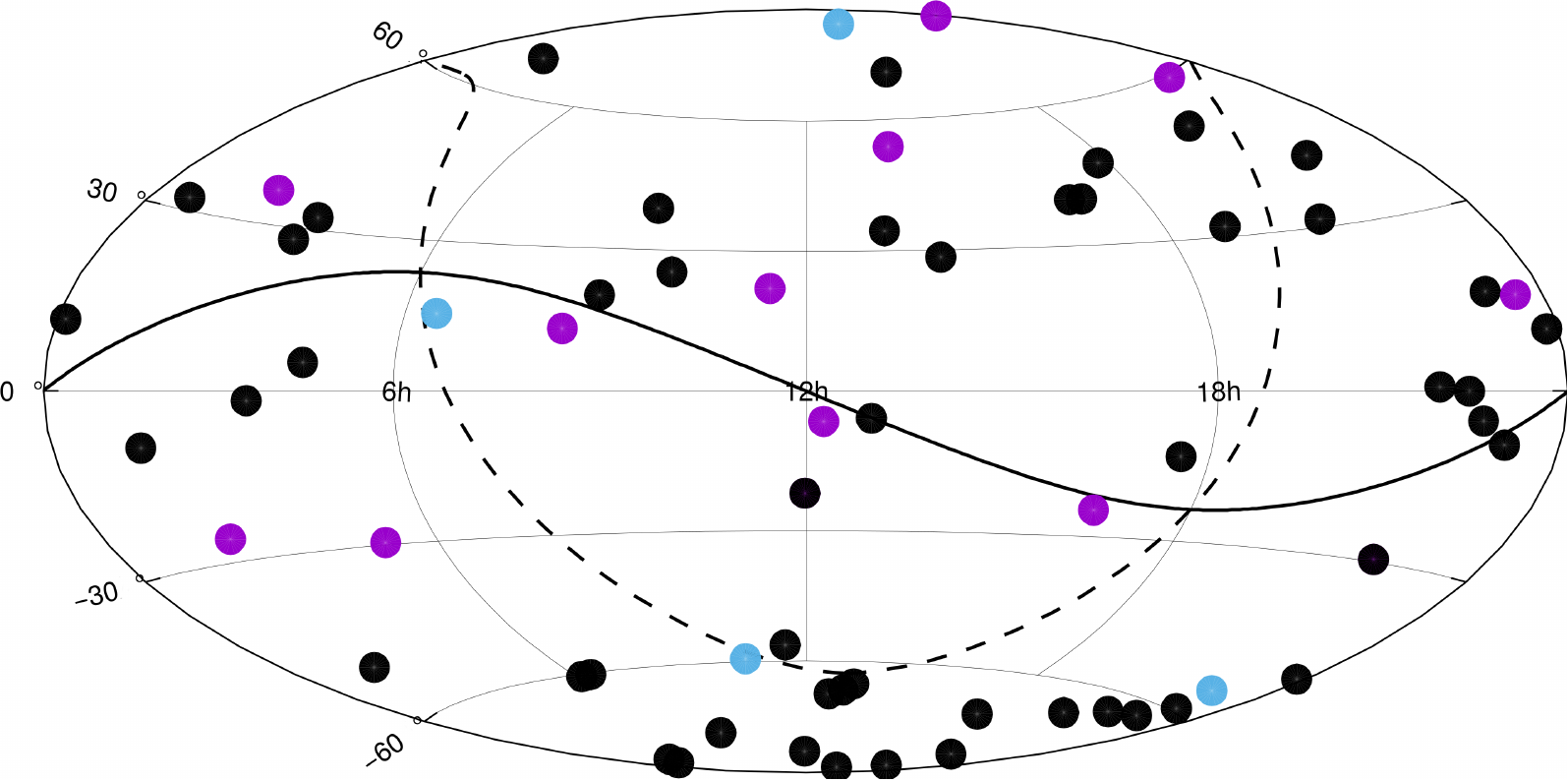}}
   \caption{Sources unique to one catalog. Light blue: GFZ, purple: K band, and black: XKa.}
   \label{fig:Unique}
\end{figure}

While the majority of radio sources exists in almost all solutions, there are a few  that are only contained in one of the catalogs (Fig. \ref{fig:Unique}). Four sources are unique to the GFZ catalog (light blue), 11 are unique to K band (purple), and 31 unique to the XKa catalog (black). It can be noted, that the far south especially profits from an extension of the SX catalog through a combination with the XKa catalog. Sixteen of the sources contained solely in the latter are south of -30$^{\circ}$ in declination, which is a sector in which SX catalogs are still more sparsely populated than in the northern hemisphere.

\subsection{Quick look data characterization}
\label{ch:quicklook}
\begin{figure*}[ht!]
        \includegraphics[width=0.98\textwidth]{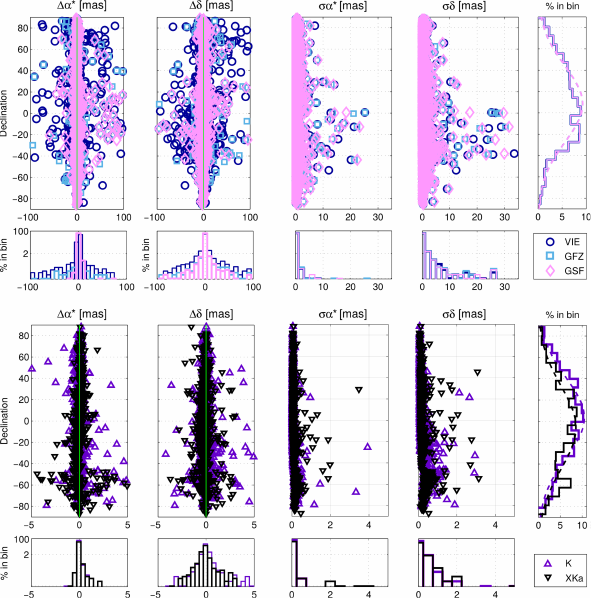} 
    \caption{Residuals w.r.t. ICRF2 (left) and standard deviations (right) vs. declination. The top row shows GSF (magenta diamonds), VIE (dark blue circles), and GFZ (light blue squares); the bottom row shows K band (purple triangles) and XKa (black triangles).}
   \label{fig:res_std_declination}
\end{figure*}

Before a combination is performed, the input data needs to be characterized in terms of their quality and possible deviations. We, therefore, first determined differences with respect to ICRF2 \citep{ICRF2}. The results are summarized in Fig. \ref{fig:res_std_declination}, depicting the residuals and respective standard deviations. The top row shows the results of the SX catalogs, while the K and XKa catalogs are shown in the bottom row. For the sake of a good visual impression, the plotting limits are set rather narrow but then exclude a very small number of sources. The truncation is of no consequence for the interpretation of the plot. The asterisk symbol ($*$) in connection with the residuals, differences, and errors of right ascension $\alpha$ denotes their scaling by declination $\delta$: $\alpha^*=\alpha \cos \delta$.

In the top row, the GFZ solution (magenta diamonds) exhibits the smallest scatter in the residuals and VIE (dark blue circles) the largest, GFZ (light blue squares) lies in between. This can also be noted in the histograms, where the logarithmic scaled y-axis gives the percentage of sources in the respective bin. For all the SX catalogs, the majority ($> 95\%$) lie in the bin $\pm$5 mas. Looking at the right ascensions of the GSF solution, the majority of the remaining sources are distributed in the adjacent bins, whereas for GFZ and VIE the decline is more gradual; the slope is considerably flatter especially for VIE. This pattern is repeated in the distribution and histogram for declination, however all catalogs show a larger scatter. Considering the standard deviations, the catalogs are remarkably similar, whether for right ascension or declination. Again, in declination the scatter is higher and the slope in the histogram flatter.

For the catalogs at higher frequencies, the residuals and the respective standard deviations scatter less. We note the reduced scale of the plots. Nevertheless, the overall patterns remain the same, hence, the majority of sources fall in the central bin ($> 80\%$ for K band, $> 90\%$ for XKa) and the slopes of the histograms are steeper for right ascension than for declination. Looking closely, a slight asymmetry can be noted in XKa (black triangles) in the residuals: for right ascension in the northern hemisphere the most of the residuals are positive and in the southern hemisphere most are negative; for declination the majority are negative. In case of the standard deviations, the declination results exhibit more scatter than in right ascension with a steady increase  in both catalogs moving south. For K band (purple triangles) the scatter in the south can be attributed to the limited coverage of the southern sky through the VLBA observations, the small number of single baseline sessions between Hartebeesthoek and Hobart, and the weak geometry of the latter. 

Similarly, the peculiar pattern in the standard deviations might be explained by the weak geometry of the network for XKa as well. The majority of the sources were observed by a three-station network with two in the northern and one in the southern hemisphere, hence making it impossible to observe sources south of -45$^{\circ}$. The sources south of that were observed using the four-station configuration, where again the southern baseline (Argentina-Australia) has limited sky coverage with respect to the northern hemisphere.

\begin{table}
\begin{small}
\caption{Statistics for the residuals w.r.t. ICRF2 of each catalog. All values are given in [mas]. The last column gives the number of sources in each solution.}
  \begin{tabular}{llrrrrrr} 
  \noalign{\smallskip}
  \noalign{\smallskip}\hline\noalign{\smallskip}
      & & mean   & $\sigma$   &   wmean   &    $\sigma_{wmean}$  &   wrms & $\#$      \\
      \noalign{\smallskip}
      \noalign{\smallskip}\hline\noalign{\smallskip}
      \multirow{2}{*}{GSF} & $\alpha*$ &  0.400   &    5.972    &   0.011     &   0.424    &   0.575 &  \multirow{2}{*}{4536}      \\
                           & $\delta$  &  0.037   &    5.157    &   -0.043    &   0.499    &   0.505 &  \\
      \noalign{\smallskip}\hline\noalign{\smallskip}                                                                                    
      \multirow{2}{*}{VIE} & $\alpha*$  &  0.472   &    10.066   &    0.010    &   1.298    &   1.377 &  \multirow{2}{*}{4521}      \\
                           & $\delta$   &  -0.010  &    10.165   &   -0.049    &   1.360    &   1.360 &  \\
      \noalign{\smallskip}\hline\noalign{\smallskip}                                                                                                                                                                        
      \multirow{2}{*}{GFZ} & $\alpha*$  & 0.321    &    6.422    &   0.016     &   0.538    &   0.619 &  \multirow{2}{*}{4537}      \\
                           & $\delta$   & 0.003    &    6.278    &   -0.085    &   0.632    &   0.639 &  \\
      \noalign{\smallskip}\hline\noalign{\smallskip}                                                                                                                                                                       
      \multirow{2}{*}{K}   & $\alpha*$  & 0.057   &    0.909    &   -0.011    &   0.452    &   0.457  &  \multirow{2}{*}{824}      \\
                           & $\delta$   & 0.002   &    1.021    &   -0.038    &   0.624    &   0.625  &  \\
      \noalign{\smallskip}\hline\noalign{\smallskip}                                                                                                                                                                        
      \multirow{2}{*}{XKa} & $\alpha*$  & 0.123   &    1.293    &    0.042    &   0.349    &   0.358  &  \multirow{2}{*}{674}      \\
                           & $\delta$   & -0.135  &     0.652   &   -0.156    &   0.395    &   0.395  &  \\
      \noalign{\smallskip}\hline\noalign{\smallskip}
 \end{tabular} 
 \label{tab:stat_cat}  
\end{small}  
\end{table}   

The statistical properties of the residuals of the individual solutions such as mean and weighted mean with the respective standard deviations and the weighted root-mean-square (WRMS) of the residuals are summarized in Tab. \ref{tab:stat_cat} and display the differences in the catalogs in numerical form. The last column gives the number of sources contained in the catalog. 

A general fact  is that the standard deviations of the (unweighted) mean and that of the weighted mean differ by an order of magnitude. This indicates that some great deviations in the data drive the unweighted mean and that a proper weighting is absolutely necessary for a valid interpretation of the data. The WRMS can be interpreted as an average value for the deviation of a single data point.

For SX, all solutions show a slight positive bias in the mean for the right ascensions with respect to ICRF2 and a negative bias for the declinations. The origin most probably resides in the ICRF2 solution with much fewer observations in the southern hemisphere. Looking at the accompanying standard deviations and WRMS, the impressions gained from Fig.~\ref{fig:res_std_declination} are confirmed. The GSF catalog shows the smallest scatter, closely followed by GFZ, whereas for VIE it is doubled. The two components of the positions have a very similar magnitude. 

An explanation for why GSF excels in these comparisons is that ICRF2 was also determined by the GSF with the Calc/Solve software. Hence, the differences between the two solution setups can be expected to be small. In the case of GFZ and VIE, a different software was used, which might result in the larger scatter. The sources that show the largest residuals with respect to ICRF2 are typically those seldomly observed and consist mainly of so-called VCS sources (VLBA Calibrator Survey; \citealt{Petrov2016} and references therein). The differences between GFZ and VIE concerning these sources point at a different handling of the VCS sessions by the individual analysts. 

The mean values and respective standard deviations for the X band and XKa catalogs reflect the higher accuracy that can be achieved at higher frequencies owing to the reduced impact of source structure group delay effects, thus the repeatabilities of the individual observations are higher. Taking the weights into account, the performance of the K band catalog is comparable to GSF. Again, the same analysis center that provided the GSF solution is also responsible for the K band catalog. 

Concluding, all SX catalogs have a similar performance, where GSF shows the smallest deviations from ICRF2 because of the reasons explained above. For this reason and for keeping the balance between the catalogs representing the three frequency bands, we only proceed with the GSF solution for the SX observations for the remainder of this publication. 

These comparisons however, made it also clear that the given standard deviations do not reflect the true accuracy of the source positions, but only the repeatability within one specific solution. For the higher frequency catalogs, the differences with respect to their a priori are overall smaller. However, distinct systematics are evident owing to weak network configurations, such as the low connectivity of the southern to the northern hemisphere in case of K band and the very weak geometry of the three-station network for XKa.

\subsection{Extended quality assessment of input to combination}

In reducing the candidates for the combination to just three input catalogs, we also should look at some other characteristics of these data. For this purpose, we applied another method to assess the a priori quality of the input to the combination. We decomposed the vector field of the positional differences of sources in two different catalogs into vector spherical harmonics \cite[VSH; e.g.,~][]{Mathews1981} of the electric (E) and magnetic (M) types, i.e., 
\begin{equation}
    \mu =  \sum\limits_{l,m} ( a_{l,m}^E Y_{l,m}^E + a_{l,m}^M Y_{l,m}^M )  \; ,
\end{equation}
where $a_{l,m}^E$ and $a_{l,m}^M$ are the coefficients and $Y_{l,m}^E$ and $Y_{l,m}^M$ and the VSH describing the given vector field $\mu$.

In principle, such a decomposition can be done to an infinite degree, where higher degrees reflect smaller details of the vector field. However, only the low degrees can be linked directly to global features. Degree one terms represent a dipole and a rotation while degree two terms describe a quadrupole. We followed the conventions introduced by \cite{Mignard2012} and the notation used by \cite{Titov2013}. 

One part of the first degree describes a rotation around the three principal axes: 
 \begin{equation}
    \begin{aligned}\label{eq:rotation}
      (\Delta \alpha~\cos~\delta)_{1,m} =&~ R_1~\cos~\alpha~\sin~\delta + R_2~\sin~\alpha~\sin~\delta - R_3~\cos~\delta \;, \\
      (\Delta \delta)_{1,m} =& -R_1~\sin~\alpha + A_2~\cos~\alpha\; ,
    \end{aligned}
\end{equation}
where $R_1=a_{1,1}^M$, $R_2=a_{1,-1}^M$, $R_3=a_{1,0}^M$. The amplitude of the rotation vector $|R|$ can be calculated with
\begin{equation}
 \label{eq:rotation_amp}
      |R|= \sqrt{R_1^2+R_2^2+R_3^2} \;.
\end{equation}

The other part describes the so-called glide toward the respective axis
\begin{equation}
    \begin{aligned}\label{eq:glide}
      (\Delta~\alpha~\cos~\delta)_{1,m} =& -D_1~\sin~\alpha + D_2~\cos~\alpha \;, \\
      (\Delta \delta)_{1,m} =&  -D_1~\cos~\alpha~\sin~\delta - D_2~\sin~\alpha~\sin~\delta + D_3~\cos~\delta \;,
    \end{aligned}
\end{equation}
where $D_1=a_{1,1}^E$, $D_2=a_{1,-1}^E$, $D_3=a_{1,0}^E$ are the components of the glide. The amplitude of the glide $|D|$ and its direction can be calculated with
\begin{equation}
 \label{eq:glide_amp_dir}
 \begin{aligned}
      |D| &= \sqrt{D_1^2+D_2^2+D_3^2} \;, \\
       D_{\alpha} &= atan~ \frac{D_2}{D_1} \;, \\
       D_{\delta} &= asin~ \frac{D_3}{|D|} \;. \\
\end{aligned}      
\end{equation}
The accompanying standard deviations can be derived using the universal propagation of uncertainty.

The quadrupolar anisotropy of the vector field of the differences is given by the development of the degree-2 VSH as follows:

 \begin{equation}
 \begin{aligned}
 (\Delta~\alpha~\cos~\delta)_{2,m} =&~a_{2,0}^{M} \sin~2\delta  \\
                             & +\sin~\delta~(a_{2,1}^{E,Re} \sin~\alpha + a_{2,1}^{E,Im} \cos~\alpha) \\
                             & -\cos~2\delta~(a_{2,1}^{M,Re} \cos~\alpha -  a_{2,1}^{M,Im} \sin~\alpha) \\
                             & -2~\cos~\delta~(a_{2,2}^{E,Re} \sin~2\alpha + a_{2,2}^{E,Im} \cos~2\alpha) \\
                             & -\sin~2\delta~(a_{2,2}^{M,Re} \cos~2\alpha - a_{2,2}^{M,Im} \sin~2\alpha) \\
 (\Delta \delta)_{2,m} =&~a_{2,0}^{E} \sin~2\delta  \\
                             & -\cos~2\delta~(a_{2,1}^{E,Re} \cos~\alpha - a_{2,1}^{E,Im} \sin~\alpha) \\
                             & -\sin~\delta~(a_{2,1}^{M,Re} \sin~\alpha +  a_{2,1}^{M,Im} \cos~\alpha) \\
                             & -\sin~2\delta~(a_{2,2}^{E,Re} \cos~2\alpha - a_{2,2}^{E,Im} \sin~2\alpha) \\
                             & +2~\cos~\delta~(a_{2,2}^{M,Re} \sin~2\alpha + a_{2,2}^{M,Im} \cos~2\alpha) \;, \\                    
 \end{aligned}
 \label{eq:quardupol}
\end{equation}

where $Re$ denotes the real and $Im$ denotes the imaginary part. All these parameters are very sensitive with respect to the choice of identical sources in both frames \cite[e.g.,][]{Mignard2018}. We used for comparisons with ICRF2 the ICRF2 defining sources \citep{ICRF2}, and for comparisons with ICRF3 the ICRF3 defining sources \citep{Charlot2019}. 

Table \ref{tab:rotParam_ind_VS_ICRF} summarizes the relative orientation and deformation parameters between the individual catalogs and ICRF2 and ICRF3, respectively. First, the very small numbers for GSF versus ICRF3 catch the eye, which is no surprise because ICRF3 and GSF are identical, save for the inflated errors in ICRF3 \citep{Charlot2019} according to Eq. \ref{eq:inflated_sig}. The very small parameters and their standard deviations also prove that the entire process of handling the normal equation systems is correct throughout. These equations are written as:
 
 \begin{equation}
 \begin{aligned}
   & \sigma^2_{\alpha*}=(1.5 \sigma_{\alpha^*})^2 + (0.03 mas)^2 \;, \\
   & \sigma^2_{\delta}=(1.5 \sigma_{delta})^2 + (0.03 mas)^2 \ .
 \end{aligned}
 \label{eq:inflated_sig}
\end{equation}

However, where the magnitude of the glide $|D|$ is reasonably small and well defined, its direction is not. Because of the small length of the vector, its direction cannot be estimated reliably, thus the large standard deviation of 140 $\mu$as in right ascension.  
In case of ICRF2, the rotation angles are reasonably small because the identical defining sources were used for their determination, however they are significant. Then again, the ICRF2 formal errors were inflated a posteriori similarly as for ICRF3, hence we can assume an overestimation of these parameters. The glide parameters, especially $D_2$ and $D_3$, are highly significant. The parameter D$_2$ describes a pattern that is similar to the effect of galactic aberration, and thus is of no surprise in case of the GSF--ICRF2 comparison, as GSF applied a correction for galactic aberration, whereas ICRF2 did not (see Chap.~\ref{ch:data}). The parameter
D$_3$ is directly connected to the source declination, as it describes a glide toward the z-axis and thus is susceptible to many modeling and analysis choices during the processing of the VLBI data and linked to the so-called declination bias \cite[e.g.,][and references therein]{Lambert2014}. This term describes an artificial displacement of the sources toward the poles and is mainly associated with the asymmetry of the VLBI network. The nature and origin of this effect are explained and explored in more detail in Chap.~\ref{ch:interComp}. 

A significant value is also found for $a_{2,0}^{E}$, which is a parameter that is connected directly to the declination of the sources as well. The amplitude of the glide is large and its direction coincides roughly with the galactic center, indicating that the main deformational difference between ICRF2 and GSF (thus also ICRF3) is due to galactic aberration. 

For K band the rotation angles with respect to ICRF2 and ICRF3 are not significant, but R$_2$ for ICRF3 is an exception to this. For the XKa catalog, R$_2$ and R$_3$ are comparably big, however these parameters are very sensitive to the choice of sources. Also, only 205 out of the 295 ICRF2 defining sources are found in the XKa catalog and 220 in the K band catalog. 

The glide parameters are more informative. For K band they are at a similar level as the GSF--ICRF2 deformations; those for XKa exceed those of GSF-ICRF2 by far pointing to the aforementioned deformations due to the weak network geometry. Looking at the higher order VSH parameters, the $a_{2,0}^{M}$ parameter for XKa stands out. While $a_{2,0}^{E}$ describes a drift toward the poles, $a_{2,0}^{M}$ describes a shearing of the two hemispheres. In this case again the network geometry manifests itself, particularly the lacking connection between north and south. The dominance of the two-baseline observations in the XKa solution has especially dramatic ramifications on the frame, which might impact negatively any combination and is investigated in Chap.~\ref{ch:results}. 

\begin{table*}
\caption{Relative orientation and deformation parameters between the original catalogs and ICRF2 and ICRF3, defining sources only. All units are $\mu$as except for D$_{\alpha}$ and D$_{\delta}$ which are in [$^{\circ}$].}
  \begin{tabular}{lrrrrrr}    
    \hline\noalign{\smallskip}  
                        & \multicolumn{2}{c}{GSF}            &\multicolumn{2}{c}{K}            &\multicolumn{2}{c}{XKa}            \\
                                        &      vs. ICRF2    &      vs. ICRF3         &  vs. ICRF2           & vs. ICRF3         &    vs. ICRF2       &   vs. ICRF3         \\
\noalign{\smallskip}\hline\noalign{\smallskip}                                                                                          
        $R_1$                             &       12 $\pm$ 7 &    -0.4 $\pm$ 0.7    &    20   $\pm$ 17         &        -2 $\pm$ 14  &   16 $\pm$ 16  &   4 $\pm$ 15    \\
        $R_2$                             &       18 $\pm$ 7 &     0.1 $\pm$ 0.7    &   -28   $\pm$ 17    &   -56 $\pm$ 14  &  -54 $\pm$ 17  & -84 $\pm$ 15    \\
        $R_3$                             &   -5 $\pm$ 6 &    -0.0 $\pm$ 0.5    &    -0   $\pm$ 10    &     3 $\pm$  8  &   47 $\pm$ 11  &  51 $\pm$ 10    \\
        $|R|$                     &       22 $\pm$ 7 &     0.4 $\pm$ 0.7    &    34   $\pm$ 17    &    56 $\pm$ 14  &   74 $\pm$ 15  &  99 $\pm$ 14    \\
        \noalign{\smallskip}\hline\noalign{\smallskip} 
        $D_1$                             &  -23 $\pm$ 7 &     0.0 $\pm$ 0.7    &   -29   $\pm$ 16    &   -11 $\pm$ 13  &  -33 $\pm$ 16  & -10 $\pm$ 15    \\
        $D_2$                             &  -75 $\pm$ 7 &     0.1 $\pm$ 0.8    &   -55   $\pm$ 15    &    27 $\pm$ 13  &  -84 $\pm$ 16  & -10 $\pm$ 14    \\
        $D_3$                             &  -98 $\pm$ 6 &    -1.5 $\pm$ 0.6    &   -64   $\pm$ 13    &    36 $\pm$ 11  & -229 $\pm$ 14  &-126 $\pm$ 12    \\
        |D|                       &  126 $\pm$ 6 &     1.5 $\pm$ 0.6    &    90   $\pm$ 14    &    46 $\pm$ 12  &  246 $\pm$ 14  & 127 $\pm$ 12    \\
        $D_{\alpha}$  [$^{\circ}$]&     -288 $\pm$ 7 &  -109   $\pm$ 140    &  -298   $\pm$ 18    &   -67 $\pm$ 33  & -291 $\pm$ 13  &-315 $\pm$ 57    \\
        $D_{\delta}$  [$^{\circ}$]&      -51 $\pm$ 2 &   -85   $\pm$ 1.8    &   -45   $\pm$  7    &    50 $\pm$  8  &  -68 $\pm$  1  & -83 $\pm$  1    \\
        \noalign{\smallskip}\hline\noalign{\smallskip}                                                                              
        $a_{2,0}^{E}   $&       61 $\pm$ 7 &   0.3 $\pm$ 0.6  &  19 $\pm$ 15      &  -38 $\pm$ 13  &   17 $\pm$ 16 &  -38 $\pm$ 15     \\
        $a_{2,0}^{M}   $&       -4 $\pm$ 8 &   0.0 $\pm$ 0.7  & -39 $\pm$ 15  &  -37 $\pm$ 12  &  293 $\pm$ 16 &  294 $\pm$ 14      \\
        $a_{2,1}^{E,Re}$&  -11 $\pm$ 9 &  -0.3 $\pm$ 0.9  & -39 $\pm$ 19  &  -42 $\pm$ 16  &  -73 $\pm$ 20 &  -68 $\pm$ 18    \\
        $a_{2,1}^{E,Im}$&        2 $\pm$ 9 &   0.9 $\pm$ 0.8  & -25 $\pm$ 19  &  -32 $\pm$ 16  &   14 $\pm$ 20 &   13 $\pm$ 18     \\
        $a_{2,1}^{M,Re}$&       -0 $\pm$ 9 &   0.1 $\pm$ 0.8  &   7 $\pm$ 17  &   13 $\pm$ 15  &   23 $\pm$ 18 &   19 $\pm$ 16     \\
        $a_{2,1}^{M,Im}$&       -6 $\pm$ 9 &   0.1 $\pm$ 0.8  & -48 $\pm$ 18  &  -42 $\pm$ 15  &  -17 $\pm$ 18 &   -7 $\pm$ 16    \\
        $a_{2,2}^{E,Re}$&       -5 $\pm$ 4 &   0.1 $\pm$ 0.3  &  -9 $\pm$ 6   &   -6 $\pm$ 5   &  -15 $\pm$ 7  &  -11 $\pm$ 6   \\
        $a_{2,2}^{E,Im}$&       -1 $\pm$ 4 &   0.1 $\pm$ 0.3  &  -5 $\pm$ 6   &   -6 $\pm$ 5   &   -7 $\pm$ 7  &   -7 $\pm$ 6   \\
        $a_{2,2}^{M,Re}$&        2 $\pm$ 4 &  -0.4 $\pm$ 0.3  &  10 $\pm$ 8   &    8 $\pm$ 6   &  -15 $\pm$ 9  &  -20 $\pm$ 7   \\
        $a_{2,2}^{M,Im}$&        3 $\pm$ 4 &   0.3 $\pm$ 0.3  &   4 $\pm$ 8   &    0 $\pm$ 6   &  -16 $\pm$ 9  &  -19 $\pm$ 7   \\
    \noalign{\smallskip}\hline\noalign{\smallskip}
\end{tabular}                       
\label{tab:rotParam_ind_VS_ICRF}                                              
\end{table*}

\section{Aspects of core shift effects}
\label{ch:angular}
An issue when combining multifrequency source positions might arise from the so-called core-shift, that is an offset between the positions in the different frequencies. This is because the apparent position of the jet base (core) in radio-loud active galactic nuclei changes with frequency due to synchrotron self-absorption. \citet{Plavin2019b} investigated 40 AGN and determined typical offsets between the core positions at 2 and 8 GHz of about 0.5~mas with a variability of the individual core positions of about 0.3~mas over ten years. At higher frequencies the core shift is less pronounced, as it follows in many cases a negative exponential distribution, leading to a core shift between X and K band of about 0.1 mas \citep{Hada2011}. However, with the assumption that these offsets have a random orientation as the jets show no preferred orientation in space and there are no global systematic differences between the positions, it only adds white noise to the position differences \citep{Lindegren2018}.

To substantiate the validity of our omission of taking into account core shift effects, we compared the positions in the individual catalogs on the basis of angular separation. We followed the approach proposed by \cite{Mignard2016} for the detection of possible core shifts between the ICRF and the Gaia-CRF (see also Chap.~\ref{ch:gaia}). 
For our test, we computed the angular separations for all common sources in the GSF solution and the other catalogs by

 \begin{equation}
 \rho=(\Delta \alpha^{*2} + \Delta \delta^2)^{1/2} \,
 \label{eq:rho}
.\end{equation}

To give more statistical meaning to the angular distances, the standard deviations can be taken into account as well. A normalized coordinate difference can be formulated as \citep{Mignard2016}

\begin{equation}
X_{\alpha^*}=\frac{\Delta \alpha^*}{\sqrt{\sigma_{\alpha^*,ref}^2+\sigma_{\alpha^*,i}^2}} \, \qquad X_{\delta}=\frac{\Delta \delta}{\sqrt{\sigma_{\delta,ref}^2+\sigma_{\delta,i}^2}}
 \label{eq:Xrade}
.\end{equation}

The normalizing factor only accounts for the statistical errors, thus does not include possible additional noise coming from a core shift. If such a core shift is present in most sources, it would be in random directions, hence contribute to the variance of the angular distances.

Taking it a step further we calculated the so-called normalized separation
\citep{Mignard2016}:

\begin{equation}
\begin{aligned}
& X_{norm}^2=\begin{bmatrix} X_{\alpha^*} & X_{\delta} \end{bmatrix} \begin{bmatrix} 1&\overline{C}\\\overline{C}&1 \end{bmatrix}    \begin{bmatrix} X_{\alpha^*} \\ X_{\delta} \end{bmatrix} \\
 \label{eq:Xnorm}
\end{aligned}
.\end{equation}

This dimensionless measure takes into account the standard deviation of the sources $\sigma$ and their correlations within the respective catalogs $C$ by introducing the quantity $\overline{C}$ as follows:

\begin{equation}
\begin{aligned}
& \overline{C}=\frac{\sigma_{\alpha*,Ref} \, \sigma_{\delta,Ref} \, C_{Ref} + \sigma_{\alpha*,i} \, \sigma_{\delta,i} \, C_{i}}{\sqrt{(\sigma^2_{\alpha*,Ref}+\sigma^2_{\alpha*,i})(\sigma_{\delta,Ref}+\sigma_{\delta,i})}}
 \label{eq:C-overline}
\end{aligned}
.\end{equation}

\begin{figure*}
\resizebox{\hsize}{!} {\includegraphics{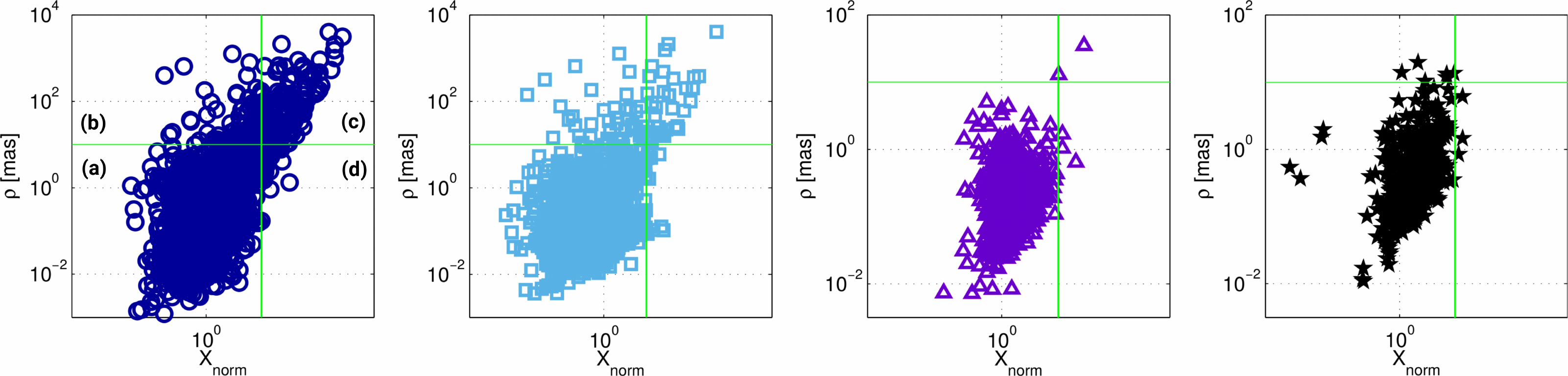}} 
\caption{Angular separation $\rho$ against the normalized separation $X_{norm}$ for \textbf{(from left to right)} VIE (dark blue), GFZ (light blue), K band (purple), and XKa (black) with respect to GSF. The vertical line is located at $X_{norm}$=4.1, the horizontal at $\rho$=10 mas.}
    \label{fig:NormVSRho}
\end{figure*}

 Following \citet{Mignard2016}, we plot $X$ against $\rho$ as shown in Fig.~\ref{fig:NormVSRho}, with a horizontal line at $\rho$ = 10 mas and a vertical line at $X$ = 4.1. These lines mark generous thresholds for the separations of the sources in terms of angular separation and  5$\%$ significance, respectively and divide the plane into the following four groups:
\begin{itemize}
\item (a): unproblematic
\item (b): not statistically significant
\item (c): most likely core shift
\item (d): moderate offsets
\end{itemize}

Looking at the resulting plots in Fig.~\ref{fig:NormVSRho} for XKa (black star cloud), no sources fall into category (c) and two sources (2018+295 and 3C119) do for K band (purple triangle
cloud), indicating a core shift. Both these sources are also included in all SX catalogs, but not in the XKa catalog. However, at closer inspection we classified both sources as outliers, as their angular distances in K band differ from the SX positions in GSF by 12.8 mas and 35 mas, respectively, which is far too big for a core shift. 

Although the choice of $\rho$ = 10 mas and at $X$ = 4.1 might be a bit arbitrary, further relaxing these thresholds does not affect the overall picture. Going below $5\%$ significance seems to be rather implausible in itself and reducing the effective distance to $5~\mu as$ would add only one additional source to the probability sample.

To emphasize the influence of the individual analysis processes, we also produced the same plots for the differences with respect to the SX catalogs VIE and GFZ. Dozens of sources fall into category (c), indicating a considerable apparent core shift with respect to the GSF positions. This however is implausible because the catalogs are determined at the same frequencies anyway. All the sources that are apparently candidates for core shifts are again VCS sources, which are sources with only few observations. This again indicates that the variances of these sources are too optimistic.

Concluding, it can be stated that at the current precision of geodetic VLBI, no core shift between the frequencies is statistically detectable. The impact of the analysis outweighs any possible positional shift, as can easily be seen in the residuals and complementary standard deviations. The differences between the individual SX catalogs are many times higher than any differences between the positions at different frequencies (see introduction of this chapter). This, in the end, also leads to the failure of any statistical testing of core shifts of a large number of sources, as the individual source position is too inaccurate at the current state and the standard deviations are too optimistic. Further, in geodetic VLBI the source position is a mean position over the entire observation period and thus cannot reflect any variations in time, which certainly occur.

\section{Combination method}
\label{ch:method}
\begin{figure}
   \centering
   \resizebox{\hsize}{!} {\includegraphics{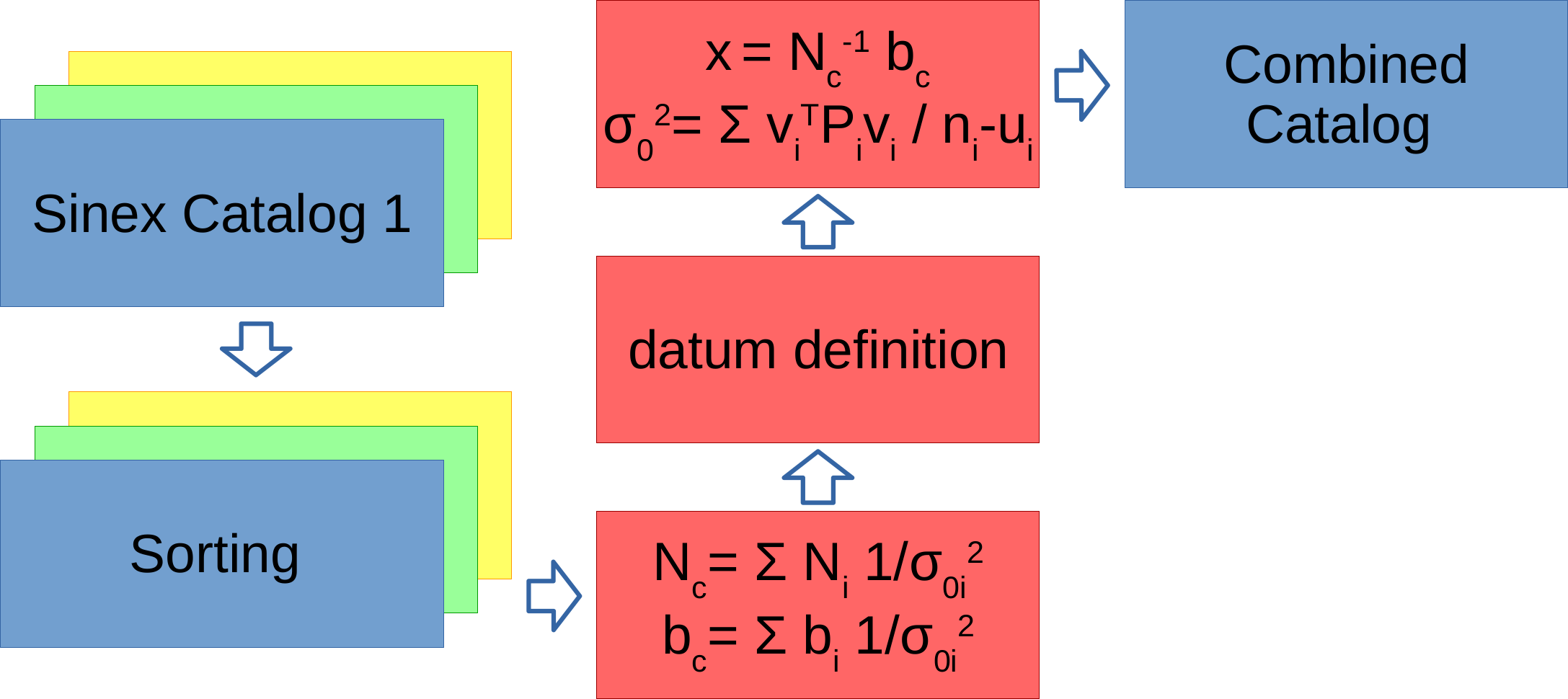}}
      \caption{Flowchart of the combination strategy.}
      \label{fig:Flow}
   \end{figure}

The combination procedure we developed, basically follows six steps as shown in the flow chart in Fig.~\ref{fig:Flow}. We denote matrices with bold capital letters and vectors with bold small letters. 

\begin{enumerate}
    \setlength\itemsep{1em}
    \item In a preparatory step to the actual combination, we perform various checks on the individual catalogs. Thus we check that the files contain the mandatory blocks and parameters to be able to calculate a solution, that the same a priori values were used, and that the information contained is free of geometric datum, that is, no orientations are defined implicitly. 
    In case of the XKa solution, which was made available to us in the form of a solution vector and a covariance matrix, we reconstruct the datum-free normal equations using the formulation given in \cite{Grafarend1985}, i.e.,
    
    \begin{equation}
        \boldsymbol{ C_{xx} = (N_{free}+B^T B) ^{-1} - B^T (B B^T B B^T )^{-1} B } \, ,
    \end{equation}
    
    where $\boldsymbol{C_{xx}}$ is the given covariance matrix, $\boldsymbol{B}$ the datum conditions (see Eq.~\ref{eq:datum}), and $\boldsymbol{N_{free}}$ the datum free normal equations. 
    
    The final test for all catalogs consists of applying the datum (NNR on all ICRF2 defining sources) and inverting the normal equation system. If all parameters are correct, the results equal the estimates reported in the SINEX files.
   
    \item In the next step, we match the NEQ of each source in the various catalogs with each other and sorted them accordingly.

    \item We rescale the NEQ accounting for the differences inherent in each catalog. The standard approach is to use the a posteriori variance factor $\sigma_0$ by multiplying each NEQ system $\boldsymbol{N}$, where
    
                \begin{equation}
                \boldsymbol{N_w}=\boldsymbol{N} \cdot 1/\sigma_0^2 \,.
                \end{equation}
                
   The a posteriori variance factor of the catalogs are 1.0289 for GSF and 0.9411 for K band. For XKa no a posteriori variance factor was available, so we set it to 2. Because initial results have shown that the XKa solution introduces significant rotations around $R_2$, we chose to down-weight this catalog. The weighting parameter of 0.05 mas$^2$ was determined empirically in such a manner that the originally dominant rotations are just no longer discernible in the residuals. This scaling can be seen as an inflation of the standard deviations of the catalog, equivalent  to the second addend in Eq.~\ref{eq:inflated_sig}.
    
    \item For the actual  combination, we follow a standard combination procedure for normal equation systems (NEQs), well known as ``Helmert blocking'' \citep{Helmert1872}, generally referred to as stacking. Basically this procedure comprises the merging of one and the same parameter contained in two or more NEQ systems into one, resulting in a combined NEQ system. Exemplarily, the formalism necessary for combining two observations is shown in this work. It can easily be expanded for more input sets along the same lines.

    We give the Jacobian matrix $A$ and the weight matrix $P$ of the combined system as
    
    \begin{equation}
    \boldsymbol{A}= \begin{bmatrix}
    \boldsymbol{A_1} \\
    \boldsymbol{A_2}
    \end{bmatrix} \,, 
    \boldsymbol{P}= \begin{bmatrix}
    \boldsymbol{P_1} & 0 \\
    \boldsymbol{P_2} & 0
    \end{bmatrix} \,.
    \end{equation}
     The corresponding NEQ system can be derived from
    \begin{equation}
    \label{eq:ATPA}
    \boldsymbol{A^T_1 P_1 A_1 + A^T_2 P_2 A_2 \cdot x} = \boldsymbol{A^T_1 P_1 l_1 + A^T P_2l_2} \,.
    \end{equation}
    
    For a two-step approach, the two systems are converted into NEQ systems $\boldsymbol{N}$ independently beforehand. However, in our case only $\boldsymbol{N}$ and the corresponding right-hand side vector $\boldsymbol{b}$ are available, which are
    
    \begin{equation}
    \label{eq:N_b}
    \begin{split}
    \boldsymbol{N_1= A^T_1 P_1 A_1} \, , \quad \boldsymbol{ b_1=A^T_1 P_1 l_1} \, ,\\
    \boldsymbol{N_2= A^T_2 P_2 A_2} \, , \quad \boldsymbol{b_2= A^T P_2l_2}\,.
    \end{split}
    \end{equation}
    
    Looking at Eq.~\ref{eq:ATPA} and Eq.~\ref{eq:N_b}, it becomes clear that for a combination of the identical parameter ${\bf x}$, the two NEQ matrices, ${\bf N}$ and the right-hand side vector ${\bf b,}$ have to be summed up as 
    
    \begin{equation}
    \begin{split}
    \boldsymbol{(N_1 + N_2) \cdot x} &= \boldsymbol{b_1+b_2} \, ,\\
    \boldsymbol{N_c \cdot x} &= \boldsymbol{b_c} \,.
    \end{split}
    \end{equation}
                            
    \item Before the combined, meaning summed-up, NEQ system can be solved, a datum has to be applied. This is done through the introduction of free-network conditions constraining the orientation of the CRF, also called NNR conditions \citep{JACOBS:2010}. 
    
    The relation of the a priori CRF and that estimated can be described by a rotation matrix $\boldsymbol{B}$ containing three rotation angles $r$. Alternatively, a parameter $d_z$ can also be introduced, which accounts for a global translation of the source coordinates in declination, reflecting systematic effects, such as the inaccuracy of the tropospheric propagation correction for sources observed at low elevations. However, because a gradient estimation is applied in the
VLBI analysis,  the parameter $d_z$ is expected to be negligible. Following \cite{Feissel2005}, the differences in coordinates for one source in the two reference frames are written as
    
        \begin{equation}
        \begin{split}
        \alpha_1 - \alpha_2 &= r_1~tan~\delta_1~\cos~\alpha_1 + r_2~tan~\delta_1 \sin~\alpha_1 - r_3 \, ,\\
        \delta_2 - \delta_1 &= -r_1~\sin~\alpha_1 + r_2~\cos~\alpha_1 + d_z \, ,
        \end{split}
        \end{equation}
        
        which translates in matrix notation to
        
        \begin{equation}
        \label{eq:datum}
        \boldsymbol{B}= \begin{bmatrix}
        & \vdots \\
        tan~\delta_1~\cos~\alpha_1 & tan~\delta_1~\sin\alpha_1 & -1 & 0 \\
        \sin~\alpha_1 &  \cos~\alpha_1 & 0 & 1\\
        & \vdots \\
        \end{bmatrix} \, ,
        \end{equation}
        with the free network parameters $\chi$:
        \begin{equation}
        \boldsymbol{\chi}= \begin{bmatrix}
        r_1 & r_2 &  r_3 & dz  
        \end{bmatrix} \,.
        \end{equation}
        Hence we obtain the extended NEQ system
        \begin{equation}
         \boldsymbol{N_c}= \begin{bmatrix}
             \boldsymbol{N_c} & \boldsymbol{B} \\
             \boldsymbol{B^T} & \boldsymbol{0}
             \end{bmatrix} \,, 
             \boldsymbol{b_c}= \begin{bmatrix}
             \boldsymbol{b_c} \\
             \boldsymbol{0} 
         \end{bmatrix} \,.
        \end{equation}
        
    \item In the last step the NEQ system is solved as
        \begin{equation}
            \boldsymbol{x=N_c^{-1} \cdot b_c~}.
        \end{equation}
\end{enumerate}

This procedure not only allows us to combine the given catalogs, but also assures the quality and correctness of the data that is to be combined.

\section{Results}
\label{ch:results}
\subsection{Solution setup}
Following the principles of the combination process for the International Terrestrial Reference Frame (ITRF; \citealp{Altamimi2016}), the first combination step would only handle the three SX solutions to produce a combined SX catalog. Since we have seen that the three solutions for the SX observations are of the same quality (Sec. \ref{ch:data}), we skipped this step in this work and selected the GSF solution as the input data set for the multifrequency combination.

Hence, we combined the NEQ of GSF and K band following the procedure discussed in the previous section to form the catalog $Combo_{KX}$. Further we added the reconstructed NEQ system of the XKa solution to form $Combo_{KXKa}$. This leads to the stacking of a total of 10720 and 12068 NEQ, respectively, resulting in combined normal matrices of 9134x9134 and 9234x9234 elements for 4567 and 4617 sources, respectively. For a datum, in both cases we chose the defining sources of the ICRF2, as these were also used for the determination of the individual catalogs. 

The complete combination results with its source position catalog and the full variance-covariance matrix are available from \url{ http://www.referenzsysteme.de} of Research Unit "Space-Time Reference Systems for Monitoring Global Change and for Precise Navigation in Space" of the German Science Foundation (FOR 1503) or at the CDS via anonymous ftp to \url{cdsarc.u-strasbg.fr} (130.79.128.5) or via \url{http://cdsweb.u-strasbg.fr/cgi-bin/qcat?J/A+A/}.

 \subsection{Intercomparisons}
 \label{ch:interComp}
As a first quality check, we look at the two combined catalogs Combo$_{KX}$ and Combo$_{KXKa}$ and compare them to ICRF2 and ICRF3. We note that ICRF3 at SX frequency and GSF are the same catalog simply represented in different ways. Only the formal errors of GSF were inflated a posteriori according to Eq. \ref{eq:inflated_sig} for the reported ICRF3 uncertainties.

The purpose of these intercomparisons is twofold. First, we want to show how the combination solutions refer to ICRF3, but also to ICRF2, in particular in terms of the low degree transformation parameters. Second, we demonstrate the quality of our combination product. Table \ref{tab:rotParam_combo} presents the orientation and deformation parameters between the respective catalogs in the same way as Tab.~\ref{tab:rotParam_ind_VS_ICRF}. Only the ICRF2 defining sources are used for the estimation of the transformation parameters with respect to ICRF2 and the ICRF3 defining sources for the estimation with respect to ICRF3. 

\begin{table}
\caption{Relative orientation and deformation parameters between the combinations $Combo_{KX}$ and $Combo_{KXKa}$ and ICRF2 and ICRF3. All units are $\mu$as except for D$_{\alpha}$ and D$_{\delta}$ which are in [$^{\circ}$].}
  \begin{tabular}{lrrrr}    
    \hline\noalign{\smallskip}  
                        & \multicolumn{2}{c}{Combo$_{KX}$}              &\multicolumn{2}{c}{Commbo$_{KXKa}$}          \\
                                        &      vs. ICRF2    &      vs. ICRF3         &  vs. ICRF2           & vs. ICRF3          \\
        \noalign{\smallskip}\hline\noalign{\smallskip}                                                                                          
    $R_1$                       &        22 $\pm$ 7  &  10   $\pm$ 1 &   19 $\pm$ 8  &    -8  $\pm$ 1     \\
        $R_2$                           &        26 $\pm$ 8  &  10   $\pm$ 1 &   19 $\pm$ 8  & -2  $\pm$ 1   \\
        $R_3$                           &       -19 $\pm$ 6  & -13   $\pm$ 1 &  -16 $\pm$ 6  &   10  $\pm$ 1    \\
        $|R|$                       &    39 $\pm$ 7  &  19   $\pm$ 1 &   31 $\pm$ 7  &  13  $\pm$ 1    \\
        \noalign{\smallskip}\hline\noalign{\smallskip} 
        $D_1$                           &       -23 $\pm$ 7  &   0   $\pm$ 1 &  -24 $\pm$ 8  &   0.3  $\pm$ 1    \\
        $D_2$                           &       -75 $\pm$ 7  &-0.2   $\pm$ 1 &  -75 $\pm$ 8  &  -0.4  $\pm$ 1    \\
        $D_3$                           &       -98 $\pm$ 6  &-0.6   $\pm$ 1 &  -97 $\pm$ 7  &  -0.5    $\pm$ 1    \\
        |D|                         &   126 $\pm$ 7  & -0.6   $\pm$ 1 &  126 $\pm$ 7  &   0.7 $\pm$ 1     \\
        $D_{\alpha}$  [$^{\circ}$]&-288 $\pm$ 7 & -279 $\pm$ 400 & -287 $\pm$ 7 & -232 $\pm$ 150 \\
        $D_{\delta}$  [$^{\circ}$]& -51 $\pm$ 2 &  -69 $\pm$ 32 &  -51 $\pm$  2 &  -47 $\pm$ 54 \\
        \noalign{\smallskip}\hline\noalign{\smallskip}                                                         
        $a_{2,0}^{E}   $&       60 $\pm$  7 & 0.3 $\pm$ 1  &   60 $\pm$ 7 & 0.2   $\pm$ 1    \\
        $a_{2,0}^{M}   $&       -4 $\pm$  8 & 0.6 $\pm$ 1  &   -5 $\pm$ 9 & 0.2   $\pm$ 1  \\
        $a_{2,1}^{E,Re}$&  -12 $\pm$ 10 &-0.2 $\pm$ 2  &  -13 $\pm$10 &-0.6   $\pm$ 2   \\
        $a_{2,1}^{E,Im}$&        4 $\pm$ 10 & 1.5 $\pm$ 1  &    4 $\pm$10 & 1.4   $\pm$ 1  \\
        $a_{2,1}^{M,Re}$&        1 $\pm$ 9  & 0.5 $\pm$ 1  &    1 $\pm$ 9 & 1.3   $\pm$ 1   \\
        $a_{2,1}^{M,Im}$&   -7 $\pm$ 9  &-0.2 $\pm$ 1  &   -7 $\pm$ 9 & 0     $\pm$ 1   \\
        $a_{2,2}^{E,Re}$&       -6 $\pm$ 4  &-0.5 $\pm$ 0.6   &   -6 $\pm$ 4 & -0.5  $\pm$ 1   \\
        $a_{2,2}^{E,Im}$&       -1 $\pm$ 4  &-0.3 $\pm$ 0.7   &   -1 $\pm$ 4 & -0.3  $\pm$ 1 \\
        $a_{2,2}^{M,Re}$&        3 $\pm$ 4  &-0.2 $\pm$ 0.6   &    3 $\pm$ 4 & -0.1  $\pm$ 1   \\
        $a_{2,2}^{M,Im}$&        5 $\pm$ 4  & 1.3 $\pm$ 0.6   &    5 $\pm$ 4 & 1.2   $\pm$ 1  \\
    \noalign{\smallskip}\hline\noalign{\smallskip}
 \end{tabular}                             
 \label{tab:rotParam_combo}                                                 
 \end{table}

We note that overall the rotations with respect to ICRF3 are about halved with respect to ICRF2, where Combo$_{KXKa}$ shows overall smaller values. Also the accompanying standard deviations are reduced significantly. This shows that the combinations are generally better aligned with ICRF3 and have a more precisely defined axis. This was expected, as the combination contains the ICRF3 equivalent GSF.

By using different subsets of sources for this transformation, we tested the stability of the axes. The scatter of the obtained rotation parameters indicate that the axes are stable to within 3 $\mu as$. 

Independently of this, both our combinations exhibit small but significant rotations with respect to both ICRFs. Whereas for Combo$_{KX}$ rotations were performed around all axis at a comparable magnitude, for Combo$_{KXKa}$ the rotation in $R_2$ is very small. This is the axis mostly affected by XKa, or rather its scaling. Without the scaling this parameter amounts to -30 $\mu$as and clearly dominates the rotations. 

In case of the drift, both combinations show a similar behavior, which means that they do not reveal any significant drifts with respect to ICRF3. This also explains the large uncertainties associated with the direction of the drift. Since the vector is very short, its direction cannot be estimated reliably. 

For ICRF2, a clear drift shows up and its direction is clearly aligned with the galactic center. This is reasonable, as all combined catalogs apply a correction for galactic aberration.
This pattern is also reflected in the VSH, specifically in the parameter $a_{2,0}^{E}$, which is by far the most significant for the comparisons with ICRF2. Also $a_{2,1}^{E,Re}$ can be identified; this parameter is linked to a shift in declination toward the poles, associated with the declination bias. In case of ICRF3, none of the VSH parameters are significant. 

Comparing the two combination products, we see that the parameters associated with deformations do not change significantly when adding XKa to the combination. This is also valid for the unscaled version, where only the rotations seem to propagate because this frame initially showed large deformations (Tab.~\ref{tab:rotParam_ind_VS_ICRF}). The reason that these do not  propagate into the combination results at all has to be seen in the combination. According to the weights, the XKa catalog is not able to map its adverse configuration onto the combination product but in itself is straightened by the Combo$_{KX}$ geometry. 

\begin{figure*}
\includegraphics[width=0.98\textwidth]{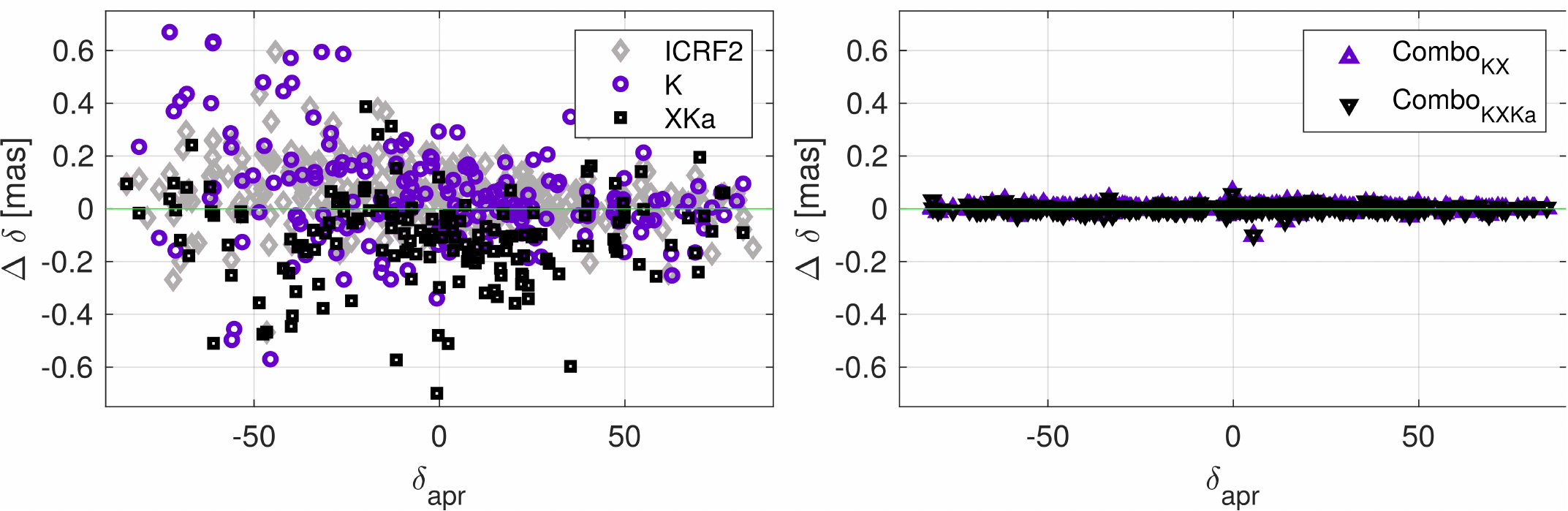}
\caption{Residuals of the defining sources plotted over declination with respect to ICRF3. \textbf{Left:} gray diamonds for ICRF2, purple circles for K band and black squares for XKa; \textbf{right:} purple triangles for Combo$_{KX}$ and black triangles for Combo$_{KXKa}$.}
    \label{fig:declination_bias} 
\end{figure*}

To investigate the propagation or mitigation of any deformations inherent in the individual catalogs to the combination we plot the residuals of the catalogs with respect to ICRF3 in declination versus the declination of the respective source. Figure \ref{fig:declination_bias} (left) shows the residuals of the individual catalogs, with respect to ICRF3 plotted versus declination, in purple and black for K band and XKa and in gray for ICRF2 (defining sources only). For each catalog, we see a larger scatter in the far south, as all observing networks have deficiencies there. K band shows no peculiar signature beyond this phenomenon. The majority of the residuals for XKa (black squares), however, form an arc between -40$^{\circ}$ and +40$^{\circ}$.  In case of ICRF2 (gray diamonds), there is also a distinct pattern of a curvature with its maximum around -40$^{\circ}$ emerges. 

This last effect is the so-called declination bias, first discovered during the determination of ICRF(1) \citep{Ma1998} and later by comparisons of ICRF2 and more recent CRF realizations (e.g., \citealp{Titov2004_CRF,Lambert2014,Liu2018}). Although the effect may have originated from different handling of tropospheric gradients, \cite{Mayer2017} and \citet{McCallumL_2017} suggested a connection to the phase calibration used at the Hobart (12~m) and Katherine (12~m) radio telescopes as main contributors.  

The situation looks very different if we plot the differences of the combination products (Fig.~\ref{fig:declination_bias}, right). The scatter is reduced significantly; no patterns, trends or other systematics are present. For both combinations most of the residuals lie between $\pm$20~$\mu$as. Thus, we can state that no declination bias is present in our combination.

\begin{figure*}
\includegraphics[width=0.98\textwidth]{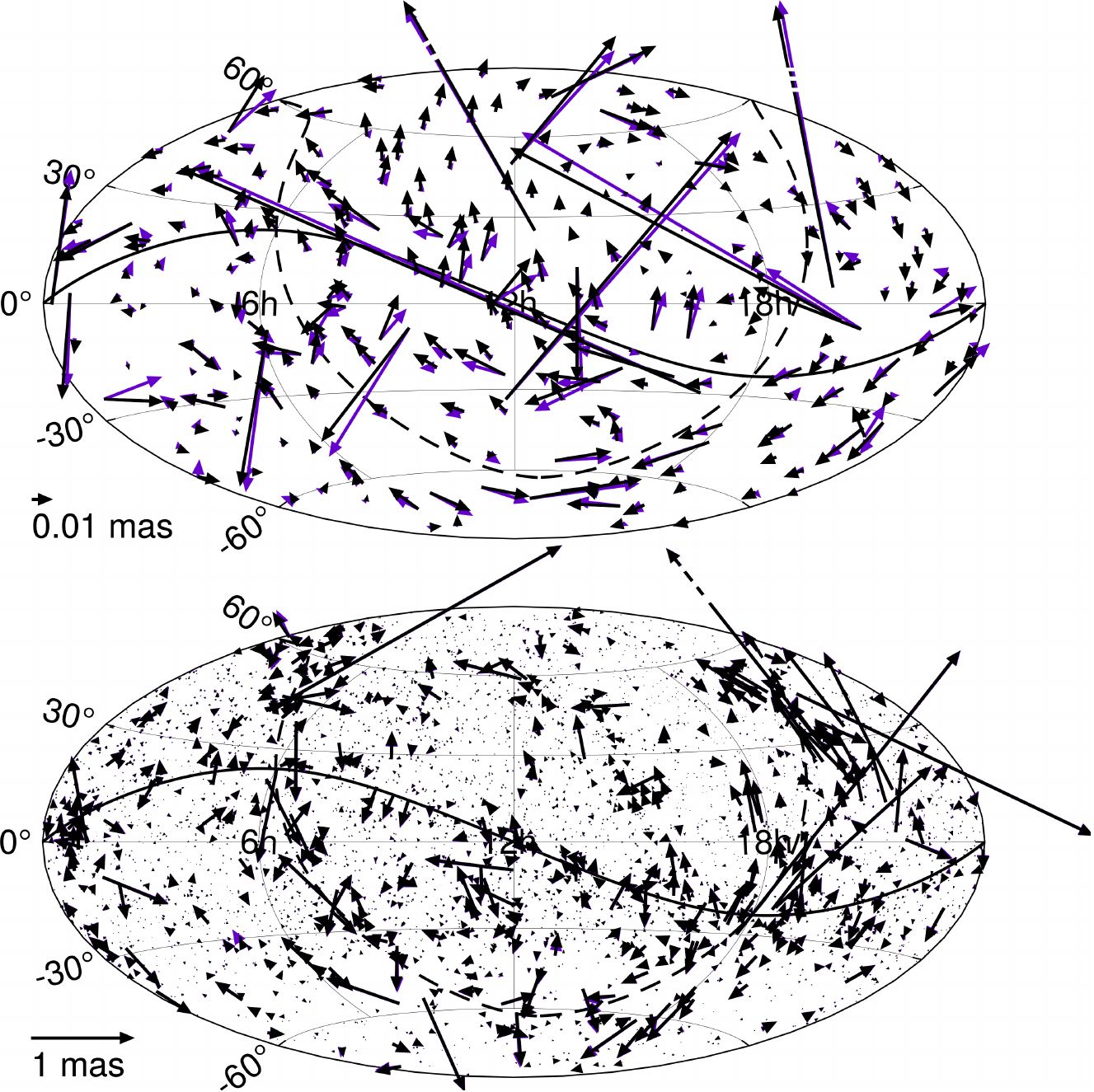} 
\caption{Residuals with respect to ICRF3, shown in purple Combo$_{KX}$ and in black Combo$_{KXKa}$. \textbf{Top:} ICRF3 defining sources only. \textbf{Bottom:} All sources. Sources with residuals >2 mas or $\sigma>$2 mas were excluded. The solid line denotes the ecliptic, the dashed line the galactic equator.}
    \label{fig:Map_res}
\end{figure*}

The celestial maps in Fig.~\ref{fig:Map_res} show the residual vectors of the combinations with respect to ICRF3: at the top for the common datum sources only ($Combo_{KX}$ in purple and $Combo_{KXKa}$ in black). At the bottom plot, all sources in both CRFs are depicted. The latter selection is restricted to those candidates with residuals and accompanying standard deviations smaller than 2 mas. This leads to approximately 4050 sources in the graph. We note the scales, which are different by two orders of magnitude in the two graphs. 

In the top plot, we can note that many of the purple vectors are covered by the black vectors or are closely aligned with them. The majority of the vectors are not more than 10~$\mu as$ in length and a faint although diffuse rotational pattern might be distinguished. Both these characteristics are in accordance with the parameters listed in Tab.~\ref{tab:rotParam_combo}. The long vectors are associated with sources for which the angular differences in the positions given in the original catalogs are large, whereas the accompanying standard deviations of the individual coordinates are small. 
In the bottom plot (note the different scale), no pattern at all can be identified, however the majority of the large differences in both combinations can be found at the observing horizon of the northern stations contributing to the SX catalog (\textasciitilde30$^{\circ}$ south) and the galactic equator where VLBI source positions are less reliable as a consequence of possible galactic extinction. 

Overall, the differences are very small and only a fraction of the differences exceed 0.5 mas. This picture is confirmed when looking at the statistics of these residuals summarized in Tab.~\ref{tab:stat_resCombo}. The last column gives the number of sources contained in the individual catalogs. 

No clear differences between the combinations can be identified, neither for the comparison with ICRF2 on the left, nor for that with ICRF3 on the right. This confirms the marginal impact of the addition of XKa on the combination, in this case more specifically on the single source positions. Again, the various mean values of Combo$_{KX}$ and Combo$_{KXKa}$ do not differ significantly. In case of ICRF2, they are at a comparable level with the corresponding values for GSF (see Tab.~\ref{tab:stat_cat}). For the comparison with ICRF3 we have smaller values for the unweighted mean. However, when taking into account the weighting, the values are larger than those with respect to ICRF2. When excluding the VCS sources, all values drop by 50\%. Looking only at the defining sources, the $wrms$ is even reduced by 80\%. That suggests that the larger deviations present between our combinations and ICRF3 are introduced mainly by the weakly observed sources in the SX catalog. 

 \begin{table*}
\caption{Statistics for the residuals of the combined solutions wrt ICRF2 and ICRF3. All values are given in [mas]. The last column gives the number of sources in each solution.}
  \begin{tabular}{llrrrrr|rrrrr|c} 
  \noalign{\smallskip}
  \noalign{\smallskip}\hline\noalign{\smallskip}
      & & \multicolumn{5}{c}{vs. ICRF2} & \multicolumn{5}{c}{vs. ICRF3}\\
      & & mean   & $\sigma$   &   wmean   &    $\sigma_{wmean}$  &   wrms &  mean   & $\sigma$   &   wmean   &    $\sigma_{wmean}$  &   wrms & $\#$      \\
      \noalign{\smallskip}
      \noalign{\smallskip}\hline\noalign{\smallskip}
      \multirow{2}{*}{Combo$_{KX}$}   & $\alpha*$ & 0.382 & 5.822 & 0.011 & 0.390 & 0.539 & 0.088 & 4.060 & -0.016 & 0.925 & 0.931 & \multirow{2}{*}{4567}      \\
                                      & $\delta$  & 0.050 & 5.117 & -0.042 & 0.492 & 0.501 & 0.062 & 4.422 & -0.018 & 1.790 & 1.792 \\
      \noalign{\smallskip}\hline\noalign{\smallskip}                                                                                                                                                                        
      \multirow{2}{*}{Combo$_{KXKa}$} & $\alpha*$ & 0.375 & 5.795 &  0.009 & 0.383 & 0.538 & 0.088 & 4.060 & -0.016 & 0.927 & 0.933 & \multirow{2}{*}{4617}      \\
                                      & $\delta$  & 0.050 & 5.089 & -0.042 & 0.482 & 0.491 & 0.062 & 4.422 & -0.018 & 1.795 & 1.797\\
      \noalign{\smallskip}\hline\noalign{\smallskip}
 \end{tabular} 
 \label{tab:stat_resCombo}     
\end{table*}                  

Concluding it can been said that our combinations are generally closer to ICRF3 than to ICRF2 as can be seen by the overall smaller deformation parameters. Any significance can only be found in the rotation parameters. We find that the axes of the combinations are stable within 3 $\mu as$.

\subsection{Precision of the combination catalogs}

\begin{figure*}
\includegraphics[width=0.98\textwidth]{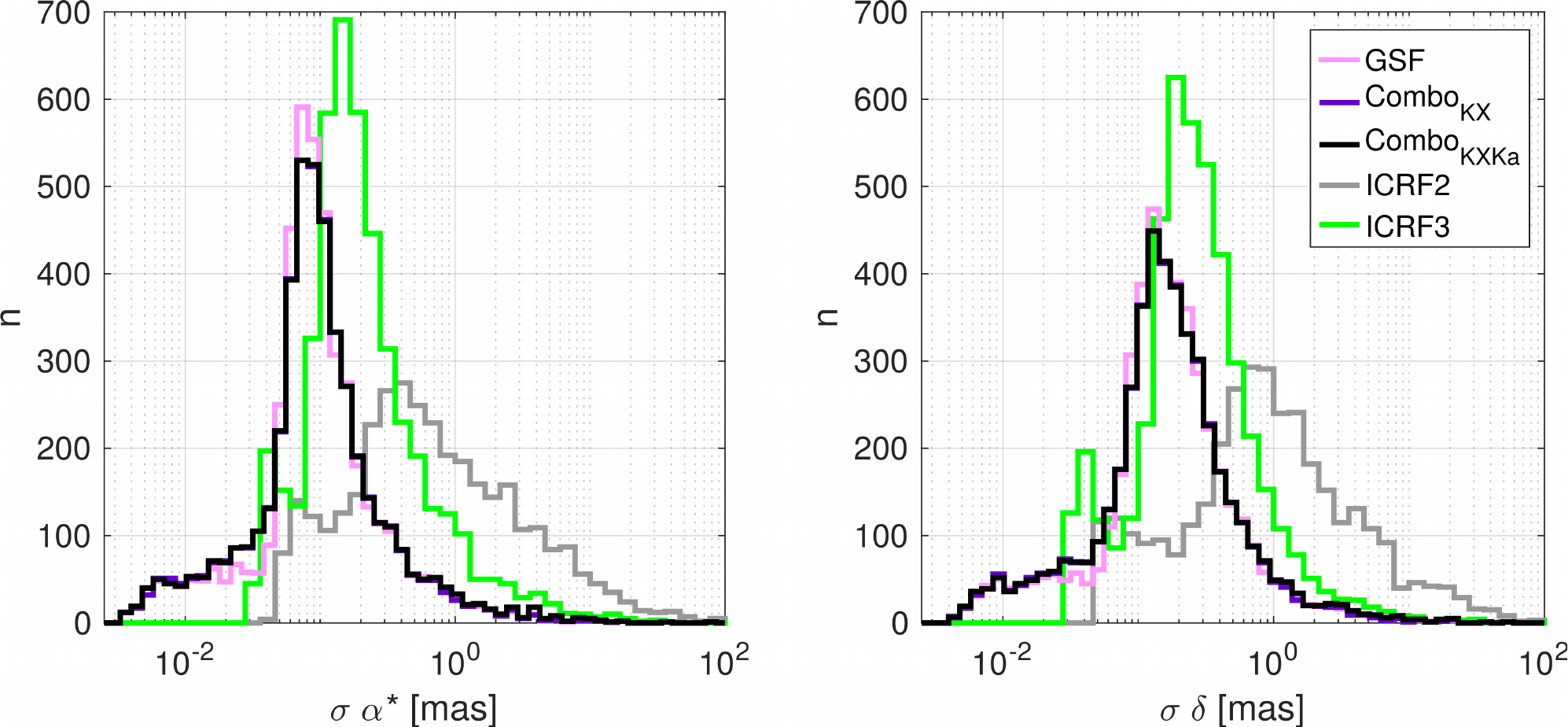} 
\caption{Distribution of the standard deviations of the estimated source positions, in pink for GSF, green for ICRF3, gray for ICRF2, and black for Combo$_{KXKa}$; on the left for right ascensions, on the right for declination. The Combo$_{KX}$ values in purple are subdued under the Combo$_{KXKa}$ graph. }
    \label{fig:disp_std}
\end{figure*}

The level of precision of the many estimated source positions and the wide range of standard deviations can best be evaluated by employing logarithmic histograms  (Fig.~\ref{fig:disp_std}). The plot on the left depicts right ascension and the plot on the right represents declination for the respective solutions. Declination shows its typically larger uncertainties. We should note that the distributions of ICRF3 (green) and ICRF2 (gray) had been scaled a posteriori. Exemplarily, this can be identified by looking at the graphs of ICRF3 (green) and GSF (pink), especially at the left-hand end of the slopes. For ICRF3 it stops abruptly around 0.02 mas, forming a peak at 0.03 mas. Within this peak most of the defining sources are found. 

The distributions of the GSF and the Combo$_{KXKa}$ solution are mostly congruent. This is valid in particular along the slope at the right-hand side. At the side of the smaller sigmas, we notice a few deviations in that the GSF counts lie below the Combo$_{KXKa}$ counts. This is then counteracted by a few more counts for GSF at the peak uncertainties. We interpret this as a change to better standard deviations for a few radio sources.

These improvements through the combination can be attributed to the fact that the input catalogs to the combinations can be considered as being uncorrelated. This originates from having entirely different observations with different observing networks and setups as well as many other differences such as the external ionosphere calibration for the K band solution. A general magnitude of the improvement cannot be quantified easily because any average sigmas of values spread over the range from two to three orders of magnitude are dominated by the large values.

 \subsection{Gaia}
 \label{ch:gaia}
Until recently there were no means to have an independent comparison and validation of a radio CRF, as all CRF, for instance galactic or optical, heavily depend on the ICRF or were not accurate enough. With the Gaia mission, this changed. Gaia is a satellite mission of the ESA launched on December 19, 2013 and is expected to operate until 2022. It is designed to scan and monitor the sky and measure positions, distances, and motions of astronomical objects in the optical regime with unprecedented precision. The final catalog is expected to consist of approximately 1 billion objects. An exhaustive description of the mission and its products can be found in the special issues of Astronomy \& Astrophysics on the Gaia data release 1 and 2 published in November 2016 and April 2018, respectively (e.g., \cite{GAIA2016}). 

Gaia Data Release 2 (Gaia DR2) contains 1.7 billion sources in the magnitude range from 3 to 21 based on observations collected during the first 22 months of its operational phase. Five astrometric parameters (positions, parallaxes, and proper motions) have been estimated for 1.3 billion sources, and approximate positions at the reference epoch J2015.5 were estimated for
an additional 361 million mostly faint sources \citep{Lindegren2018}. 

This was achieved in a two step approach. In a primary solution, the satellite attitude and astrometric calibration parameters were obtained using selected sources, amounting to 1$\%$ of the input data. This primary solution was aligned with the ICRS by means of 2844 quasars that were also visible in the optical frequency range and had an ICRF3 counterpart. The selection criteria and theoretical background are described in \cite{Lindegren2012}. In the second step, the five astrometric parameters of every source are adjusted using fixed attitude, calibration, and global parameters from the preceding primary solution.

The resulting catalog contains 556,869 quasars, 2820 of which have a counterpart in the ICRF. These quasars realize the nonrotating global reference frame called Gaia-CRF2 \citep{Mignard2018}. The agreement of the Gaia radio positions and the  ICRF3 prototype solution, which was used for this procedure, was reported to be at the level of 20 to 30 $\mu$as. For the first Gaia-CRF about 6$\%$ of the sources were found to have a significant offset \citep{Mignard2016,Petrov2017}. With the second, more accurate Gaia data release the majority of these offsets were linked to the presence of optical structure because most offsets occur along the jet \citep{Mignard2018,Plavin2019,Petrov2019}.

For our comparison of the individual and combined catalogs with the Gaia-CRF2, we again made use of the transformation parameters introduced earlier. In this context, we followed the recommendation of \citet{Petrov2019} and rescaled the uncertainties by 1.3 for the VLBI frames and by 1.06 for the Gaia-CRF2. Additionally, as the determination of the transformation parameters is very sensitive to the selection of sources, we chose to use a similar outlier detection algorithm as described in \citet{Mayer_diss}, which left us with 2330 sources out of the original 2820.

Table~\ref{tab:rotParam_Gaia} summarizes the resulting orientation and deformation parameters between all our catalogs and Gaia-CRF2. The rather big rotations between GSF representing ICRF3 and Gaia most likely originate from the fact that the prototype ICRF3, which Gaia used for its alignment, only encompasses data until November 2017 and does not exactly correspond to the final ICRF3 in terms of data input. Also, in contrast to the final ICRF3, the prototype solutions included no correction for galactic abberation. The largest impact, however, which is valid for all catalogs, has the choice of sources that were used for the original alignment of Gaia with the ICRS and the choice of sources for the comparisons made by us. Where in the primary solution 2844 sources were used to orient Gaia-CRF2, in the secondary 2820 were used (see above). Among these are sources with angular separations of many milliarcseconds, many of which are VCS sources. These sources are very sparsely observed in SX, thus have an unreliable accuracy and are therefore never used for the definition of the datum in VLBI. Most of these large outliers were eliminated through our selection procedure. We finally used 2330 sources for the transformation between Gaia-CRF2 and GSF and 511 and 431 sources for K band and XKa, respectively. The findings of \citet{Mignard2018} comparing Gaia-CRF2 and the ICRF3 prototype solution confirm that large rotational and deformational parameters are the result of the different source selections. In this context, it is advisable to focus more on the relative differences of the parameters than their absolute values. It should be noted that the combinations show rotations comparable to GSF; the total magnitude however is considerably smaller for Combo$_{KXKa}$. 

Ignoring the rotation parameters in Tab.~\ref{tab:rotParam_Gaia} due to the reasons explained, we can see that significant glide parameters are found in $D_2$ for K and $D_3$ for XKa, representing a glide in y and z, respectively. In case of the combinations, all glide parameters are above significance: on one hand the magnitude lies well below the K and XKa catalogs, on the other hand  it is twice as large as that of GSF. The direction of the total glide, however, does not align with the galactic center as was often the case before. Hence, the galactic aberration can be ruled out as a possible source, but besides that no conclusion can be drawn. 

For the VSH, the largest value is found for XKa in $a_{2,0}^{M}$ which describes a sharing of the two hemispheres. This parameter is also significant for K band, which shows additionally a significant drift toward the poles ($a_{2,0}^{M}$), as does GSF. This feature seems to be carried over to the combinations, as they show significant $a_{2,0}^{M}$ values as well. Generally the quadrupole deformations seem to be more pronounced compared to the previous test, yet the uncertainties increased as well. This indicates more complex, although less distinct, deformations. However, it can be noted that the  deformations of the combined catalogs are of a comparable level with those of GSF. 

Finally, we also want to note that at this point it cannot be excluded that uncorrected systematic effects are still present in Gaia-CRF2. Thus, together with the ambiguous alignment of it with the ICRS, any definitive interpretation are impeded. The next Gaia release Gaia-DR3, expected at the end of 2020, will help to clarify this point.

\begin{table*}
\caption{Relative orientation and deformation parameters between the individual catalogs and Gaia. All units are $\mu$as except for D$_{\alpha}$ and D$_{\delta}$, which are in [$^{\circ}$].}
  \begin{tabular}{lrrrrrr}    
    \hline\noalign{\smallskip}  
        vs. Gaia        &      GSF          &      K     &   XKa          &   Combo$_{KX}$      &    Combo$_{KXKa}$   \\
        \noalign{\smallskip}\hline\noalign{\smallskip}                                                                                           
        $R_1$                           &  -21 $\pm$ 13  & -44 $\pm$ 22  &  -8 $\pm$ 23  &       -22 $\pm$ 9  &  -25 $\pm$ 9 \\
        $R_2$                           &       75 $\pm$ 12  & 108 $\pm$ 22  & -26 $\pm$ 23  &   51 $\pm$ 9  &   45 $\pm$ 9   \\
        $R_3$                           &        9 $\pm$ 11  &  16 $\pm$ 15  &  27 $\pm$ 18  &  -14 $\pm$ 8  &  -11 $\pm$ 8 \\
        $|R|$                   &       79 $\pm$ 12  & 118 $\pm$ 22  &  38 $\pm$ 20      &  57  $\pm$ 9  &   53 $\pm$ 9   \\
        \noalign{\smallskip}\hline\noalign{\smallskip} 
        $D_1$                           &       -7 $\pm$ 13  & -23 $\pm$ 21  & -22 $\pm$ 23  &  -25 $\pm$ 9  &  -25 $\pm$ 9  \\
        $D_2$                           &        0 $\pm$ 12  &  82 $\pm$ 21  &   8 $\pm$ 22  &   12 $\pm$ 8  &   12 $\pm$ 8  \\
        $D_3$                           &   13 $\pm$ 12  &  36 $\pm$ 18  &-113 $\pm$ 19  &   21 $\pm$ 8  &   21 $\pm$ 8   \\
        |D|                     &       15 $\pm$ 12  &  93 $\pm$ 20  & 116 $\pm$ 20      &   35 $\pm$ 9  &   35 $\pm$ 8   \\
        $D_{\alpha}$  [$^{\circ}$]& -2 $\pm$  4  & -74 $\pm$ 19  & -20 $\pm$ 26  & -24 $\pm$ 10   & -25 $\pm$ 11 \\
        $D_{\delta}$  [$^{\circ}$]&     61 $\pm$ 21  &  23 $\pm$ 10  & -77 $\pm$ 2   &  37 $\pm$ 11   &  37 $\pm$ 10 \\
        \noalign{\smallskip}\hline\noalign{\smallskip}                                                                               
        $a_{2,0}^{E}   $&       40 $\pm$ 13 &  54 $\pm$ 22  &     0 $\pm$ 23  &   53 $\pm$ 9  &   53 $\pm$ 9 \\
        $a_{2,0}^{M}   $&       19 $\pm$ 13 & -49 $\pm$ 20  &   302 $\pm$ 23  &   27 $\pm$ 8  &   26 $\pm$ 8 \\
        $a_{2,1}^{E,Re}$&        7 $\pm$ 15 & -34 $\pm$ 24  &   -90 $\pm$ 28  &   14 $\pm$ 10 &   15 $\pm$ 10 \\
        $a_{2,1}^{E,Im}$&  -13 $\pm$ 15 & -28 $\pm$ 24  &    67 $\pm$ 28  &    7 $\pm$ 10 &    7 $\pm$ 10 \\
        $a_{2,1}^{M,Re}$&       15 $\pm$ 14 & 117 $\pm$ 23  &    61 $\pm$ 25  &   33 $\pm$ 10 &   32 $\pm$ 10 \\
        $a_{2,1}^{M,Im}$&       33 $\pm$ 15 &  44 $\pm$ 25  &    68 $\pm$ 27  &  -10 $\pm$ 10 &  -10 $\pm$ 10 \\
        $a_{2,2}^{E,Re}$&   -0 $\pm$ 7  & -78 $\pm$  9  &    17 $\pm$ 11  &   -2 $\pm$  5 &    2 $\pm$ 5 \\
        $a_{2,2}^{E,Im}$&        8 $\pm$ 7  & -18 $\pm$ 10  &    -6 $\pm$ 11  &   -4 $\pm$  5 &    4 $\pm$ 5 \\
        $a_{2,2}^{M,Re}$&        0 $\pm$ 7  &  23 $\pm$ 11  &     5 $\pm$ 12  &   18 $\pm$  5 &   17 $\pm$ 5 \\
        $a_{2,2}^{M,Im}$&  -14 $\pm$ 7  & -14 $\pm$ 11  &     0 $\pm$ 12  &   -3 $\pm$  5 &   -3 $\pm$ 5 \\
    \noalign{\smallskip}\hline\noalign{\smallskip}
\end{tabular}
 \label{tab:rotParam_Gaia}                                              
 \end{table*} 
 
 Summarizing, these comparisons have shown that the estimated transformation parameters are heavily dependent on the choice of the sources selected for this process. The transformation parameters and their standard deviations are smallest for the GSF SX catalog (Tab. \ref{tab:rotParam_Gaia}). However, they do not vanish as we would expect owing to the alignment of the Gaia catalog with ICRF3. The reason is that it was an ICRF3 prototype catalog and, in particular, it is intransparent which sources were used for the alignment. Also, among the large set of sources used, many are sparsely observed in SX, thus having a questionable accuracy. Through this, the alignment with the ICRF3 might be compromised. 
Consequently, also the K and XKa band catalogs in Tab. \ref{tab:rotParam_Gaia} have slightly larger values for the rotations and the glide parameters because they were aligned to ICRF3 with some other selection of sources. 
The combined catalogs, in turn, show values comparable to GSF for all of parameters with standard deviations reduced by a factor of about $\sqrt{2}$. We suspect that the combination has a minute impact on the positions of those sources which are present in all three catalogs, thus, then also affecting the transformation parameters. In any case, it should be emphasized that the final combination catalog $Combo_{KXKa}$ does not show any pathological rotations or deformations of the frame. Rather the opposite is the case. The combined frame maintains a close alignment not only to ICRF3 but also to the Gaia-CRF2.

\section{Conclusions and outlook}
\label{ch:conclusions}
In this paper, we have presented a method to combine independent, multifrequency radio source catalogs consistently. We use datum free normal equation systems of the least-squares adjustment process instead of position catalogs with their standard deviations. The process itself primarily consists of Helmert stacking of the equation systems. The primary advantage of this approach is that the full covariance information across all position components of all radio sources is carried over and available for the combined catalog. This is important for a proper interpretation of the results and their statistics.

Initially, three solutions were provided for the observations at SX band and one solution was provided each for K band and XKa band observations by five different analysis centers. In a preparatory step, we looked at possible offsets between the source positions of different analysis centers and at different frequencies. For this, we have to keep in mind that we are referring to the mean of the source positions over their entire observation histories as was applied for the ICRF3 computations. We found that the differences between the three SX catalogs exceed any differences between SX and its counterparts at a higher frequency. We conclude that the current accuracy of geodetic VLBI does not suffice to detect possible core shifts. Hence, we see no impediment to combining position catalogs of different radio frequencies. 

The result of the combination is a three-frequency CRF containing precise positions of 4617 compact radio astronomical objects. Of these, 4536 were measured at 8~GHz, 824 sources were observed at 24~GHz, and 674 at 32~GHz, leading to a frequency overlap of most of the latter two groups of sources. Only 11 sources were unique to the K band catalog and 31 to the Ka band catalog. The latter have a considerable impact on the distribution of sources in the deep south. All of these are now incorporated into the combined catalog in a rigorous fashion. This means that the full covariance information is transferred to the combined catalog with the effect that now the covariance information is available across all radio sources.

All those sources, which have appeared in more than one catalog, are virtually "redetermined" on the basis of an increased number of observations in the stacking process. It should be emphasized that through proper weighting of the input data, no geometric deformation of the final catalog is caused by network effects identified in the input catalogs. Although this, for now, excludes the issues of core shift, the standard deviations for some of the sources have improved significantly. With this, the network deficiencies were also mitigated.

In addition to that, we assessed the quality of the combinations with various comparisons on the basis of estimated transformation parameters. The rotation and deformation parameters with respect to ICRF2 and ICRF3 remain within reasonable bounds. The frame is aligned with ICRF2 within $\pm$3 $\mu$as with an average positional uncertainty of 0.1 mas in right ascension and declination. The alignment with ICRF3 proves to be better in terms of alignment and shows considerably less deformations. 

A crucial point in terms of alignment and deformations, however, proved to be the choice of identical sources, especially in case of the comparisons with Gaia-CRF2 positions. Considering this issue, no definitive statements with respect to the Gaia-CRF2s are possible at this stage, as too many unknown factors remain. However, we find real differences between source positions in the radio frequency catalogs and the Gaia-CRF2  frame. Their origins have to be investigated further to determine if they are an issue inherent in one or all VLBI frames or in the Gaia reference frame. Overall we conclude that the combination benefits any comparison with Gaia.

Finally, it can be stated that the tools for producing a CRF from VLBI observations as a combination solution based on normal equation systems and with full covariance transfer have been developed and are available. This can be employed for any CRF catalog combination but also for the next realization of the ICRS if this is a radio reference frame again.

\begin{acknowledgements}
Special thanks go to all the contributors of catalogs and their supporting organizations, namely (in alphabetical order) David Gordon (NASA-GSFC), Chris Jacobs (NASA-JPL), David Mayer (former TU Vienna, now BEV), Tobias Nilsson (former GFZ, now Lantm\"{a}teriet), Elena Skurikhina (IAA), Oleg Titov (GA), and all members of the International VLBI Service for Geodesy and Astrometry. The authors thank the German Research Foundation (Deutsche Forschungsgemeinschaft, DFG) for its financial support (NO318/13-1).
This work has made use of data from the European Space Agency (ESA) mission {\it Gaia} (\url{https://www.cosmos.esa.int/gaia}), processed by the {\it Gaia} Data Processing and Analysis Consortium (DPAC,
\url{https://www.cosmos.esa.int/web/gaia/dpac/consortium}). Funding for the DPAC has been provided by national institutions, in particular the institutions participating in the {\it Gaia} Multilateral Agreement.
We also thank the anonymous reviewer for the constructive critique and comments that helped to improve the article.

\end{acknowledgements}

\bibliographystyle{plainnat}  
\bibliography{mybib} 

\end{document}